# A method for decompilation of AMD GCN kernels to OpenCL

*K. I. Mihajlenko*[a,b], *Master Student, Junior Programmer, orcid.org/0000-0002-6168-2653, Kristina.Mihajlenko@gmail.com*
*M. A. Lukin*[a,b], *PhD, Tech., CTO, orcid.org/0000-0002-1088-3324, lukinma@gmail.com*
*A. S. Stankevich*[a], *PhD, Tech., Associate Professor, orcid.org/0000-0002-3532-8941, stankev@itmo.ru*
[a]ITMO University, 49, Kronverkskii Pr., 197101, Saint-Petersburg, Russian Federation
[b]Sudo Ltd., 20, Nahimov St., 199226, Saint-Petersburg, Russian Federation

***Introduction:*** Decompilers are useful tools for software analysis and support in the absence of source code. They are available for many hardware architectures and programming languages. However, none of the existing decompilers support modern AMD GPU architectures such as AMD GCN and RDNA. ***Purpose:*** We aim at developing the first assembly decompiler tool for a modern AMD GPU architecture that generates code in the OpenCL language, which is widely used for programming GPGPUs. ***Results:*** We developed the algorithms for the following operations: preprocessing assembly code, searching data accesses, extracting system values, decompiling arithmetic operations and recovering data types. We also developed templates for decompilation of branching operations. ***Practical relevance:*** We implemented the presented algorithms in Python as a tool called OpenCLDecompiler, which supports a large subset of AMD GCN instructions. This tool automatically converts disassembled GPGPU code into the equivalent OpenCL code, which reduces the effort required to analyze assembly code.

## Introduction

OpenCL [1] is a widespread standard for high performance computing. It is supported by all of the modern graphics processing unit (GPU) and central processing unit (CPU) vendors in contrast to vendor locked Compute Unified Device Architecture (CUDA) [2]. In particular, both Nvidia and AMD GPU support OpenCL. There are great general-purpose computing on graphics processing units (GPGPU) development tools in the Nvidia ecosystem, but AMD development tools have fallen behind the Nvidia ecosystem. Sometimes developers need to analyze assembly code for implementing better optimizations or reverse engineering. Nevertheless, there is no public decompilation tool for AMD GPU assembly. Decompiler also allows supporting programs without source code and checking for undocumented functions and backdoors. [3, 4]

OpenCL is designed to unleash the power of massively parallel processors. The OpenCL platform consist of a *host* (typically a CPU) and a set of *compute devices* (or, simply, *devices*). In this paper, devices are AMD GPUs. To avoid confusion, we denote by *program* the code executed on the host and by *kernel*, the code executed on the device. Each compute device consists of a set of *compute units*. Each compute unit consists of a set of *processing elements*.

Massive parallelism means a large number of launched processes. The process index space could be one-, two-, or three-dimensional. The set of launched process indices is called *NDRange* [5]. NDRange is divided by equal-sized *work-groups* (Fig. 1). NDRange size must be divisible by work-group size on each dimension. Otherwise NDRange size is automatically increased on each dimension to fulfill this requirement. The single process is called *work-item*. Each work-item has a unique identifier (ID) in NDRange index space (*global id*) and a unique ID in its work-group (*local id*). Each work-group also has a unique ID (*work-group ID*).

OpenCL defines four types of memory:
— global memory — a memory accessible to read and write to host and all work-items in the NDRange space;
— constant memory — a region of host-allocated global memory that is not changed during kernel execution;
— local memory — a memory accessible to work-items in a single work-group;
— private memory — a memory accessible to a single work-item.

## The AMD GCN architecture

The AMD GCN architecture [6] is related to OpenCL platform model. A GPU device consists of several compute units. Each compute unit has four *single instruction, multiple data* (*SIMD*) *Vector Units* for computing and one *SIMD Scalar Unit* for flow control. Each SIMD Unit has 16 *processing elements*. One processing element contains one arithmetic logic unit (ALU) and can execute a single OpenCL work-item. Thus, one

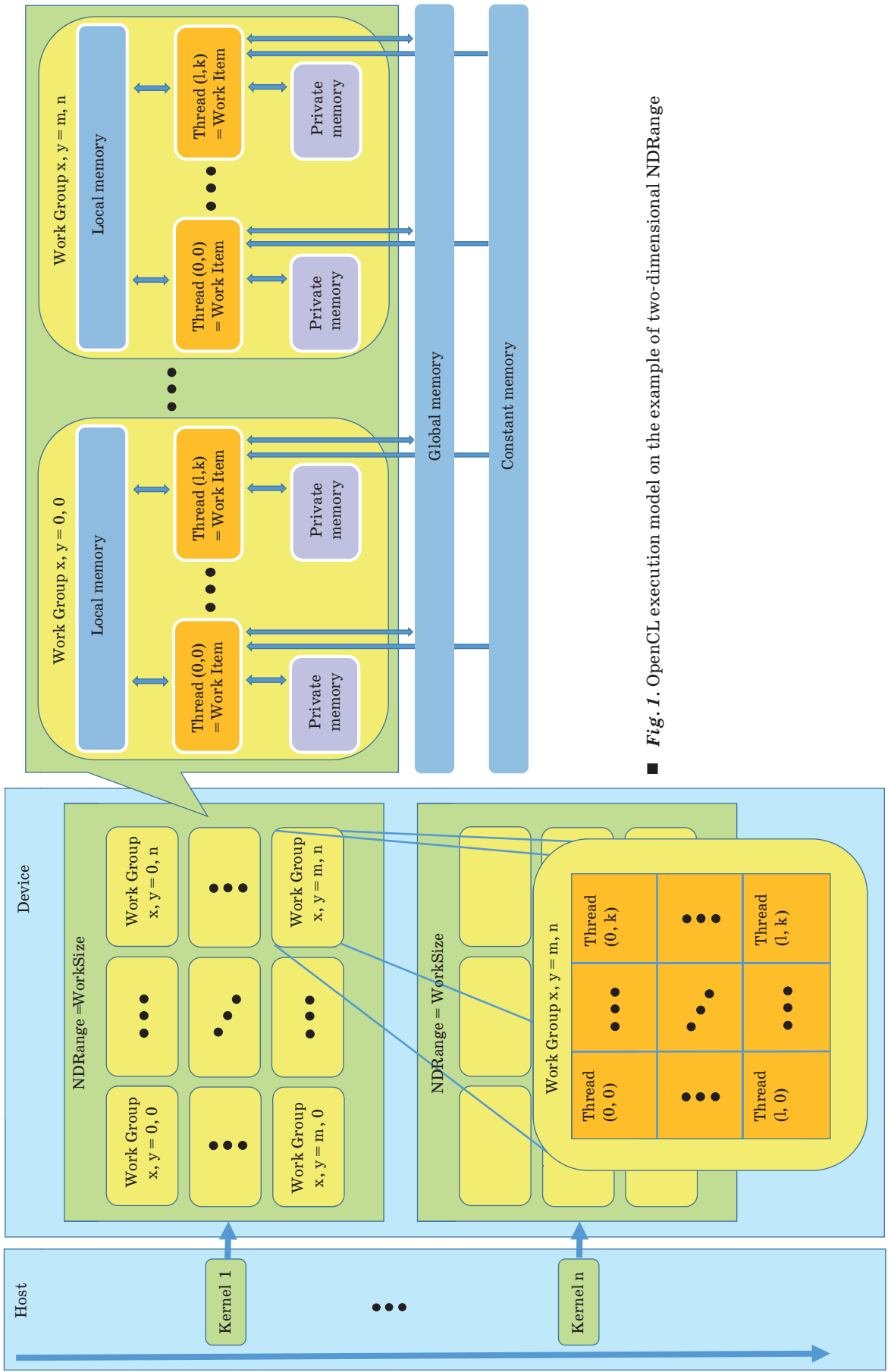

■ *Fig. 1.* OpenCL execution model on the example of two-dimensional NDRange compute unit contains 64 ALU. Compute units work independently.

GCN devices have two-level data cache hierarchy. Each compute unit has L1 data cache and an entire GPU device has L2 data cache. Also they have 32 KiB instruction cache. If kernel does not fit in instruction cache, it has significant performance decrease. This fact encourages GPGPU developers to decompose a compute task between small kernels.

AMD GCN devices have an equivalent to each type of OpenCL memory. Global and constant memories from OpenCL are represented by *video random access memory* (VRAM). The equivalent of OpenCL local memory is *Local Data Share* (LDS). Data access in LDS is orders of magnitude faster than that of a VRAM. Private memory is stored in registers. Data access in registers is orders of magnitude faster than LDS. If there are not enough registers, a region in VRAM is allocated for private memory. These additional "registers" are named *scratch registers*. Usually scratch registers are in data cache and decrease performance by not really much. Registers are 32-bit but they can be combined into pairs for 64-bit instructions. Registers are the most expensive and valuable memory resource. Each work item can have at most 256 vector registers (VGPR) and 104 scalar registers (SGPR). Moreover, a compute unit has only 2048 registers for 64 ALU.

The lowest group of work-items that flow control can affect is named *wavefront*. This means that all the work-items in a single wavefront have the same program counter. All the work-items in a wavefront execute all branching paths (with the exception of a case when all the work-items choose the same conditional jump). Irrelevant branch paths are executed without any effect. Each SIMD Vector Unit can run from one to ten wavefronts depending on the used VGPRs, SGPRs and LDS.

AMD GCN has two different application binary interfaces (ABI) [7]. The first one comes with Windows Adrenaline or Linux AMDGPU-Pro driver. The second one comes with Linux-only ROCm driver. In this paper the first one is considered. ABI defines data and kernel parameters' location in memory. Some parameters are stored in registers, others are in VRAM. More detailed location of parameters will be considered in the next sections.

## Statement of the problem

The purpose of this work is to create a decompiler for GCN assembly. It takes a disassembled file as input and translates it into its equivalent in OpenCL. Since there are no OpenCL decompilers for AMD GPUs, the following state-of-the-art theoretical [8–18] and instrumental [19, 20] solutions for C and C++ were considered as a basis:

— Ida Pro (Hex-Rays plugin): Intel x86 / x64, ARM;
— GHIDRA: Intel x86, ARM, AVR, MIPS, PIC, PowerPC;
— RetDec: Intel x86 / x64, ARM, MIPS, PIC32, PowerPC;
— Hopper: Intel x86 / x64, ARM, PowerPC;
— Snowman: Intel x86, AMD64, ARM.

As a result of research to achieve this goal, the following tasks were formulated:

1) extraction of the body of the program and data of the CPU module;
2) search for memory accesses;
3) search for control structures;
4) data type recovery.

The result of solving these tasks is a translation assembly code to an OpenCL code. Out method consists of the following steps:

1. Separation of program body, configuration part and kernel name.
2. Initialization of registers and kernel parameters using application binary interface.
3. Assembly instructions processing: control flow graph construction and determination of stored in registers data types.
4. Transformation control flow graph into region graph and its further processing (determination of flow-control instructions).
5. OpenCL code generation using processed region graph.

## The body extraction

Extracting the body of the program is a small, but quite important task, serving as a preparatory stage for further decompilation. In addition, here we parse config section with work group size, number of index range dimensions, and other kernel properties. An example of the structure of the program body is shown in Listing 1.

*Listing 1.* An example of the structure of the program body

```
.kernel [kernel name]
    .config
        dims xyz
        .cws 8, 8, 2
            [other kernel configuration]
    .text
        [program body]
        s_endpgm <- end of program
```

This config means 3D index range with workgroup size $8 \times 8 \times 2$ (128 threads in total).

The algorithm of body extraction is presented in listing 2.

*Listing 2.* The algorithm of body extraction

```
parse_status = "start"
instruction_set = []
config_set = []
program_name = ""
for current_row in bode_of_file:
  if current_row contains ".kernel"
    if parse_status == "instruction":
      parse_status = "kernel"
      process_data(program_name, config_set, instruction_set)
      instruction_set = []
      config_set = []
    program_name = current_row.split()[1] // take the second word.
  if current_row == ".config":
    parse_status = "config
  elif current_row == ".text":
    parse_status = "instruction"
  elif current_row == "instruction":
    instruction_set += current_row
  elif current_row == "config":
    config_set += current_row
  else:
    continue
process_data(program_name, config_set, instruction_set)
```

The program body consists of a sequence of assembly instructions. Most of GCN assembly instruction names consist of three parts delimited by symbol "_". In this paper they are called prefix, root and suffix.

Prefix means one of the following instruction types:

— Scalar instructions. Operands are mostly SGPRs. These instructions are used to control flow instructions, VRAM access, thread synchronization, atomic operations and others. The prefix is "s".

— Vector instructions. Operands are mostly VGPRs. These instructions are used for computing. The prefix is "v".

— Data share operations. Instructions for manipulating with LDS. The prefix is "ds".

— FLAT instructions. Operands are mostly pairs of VGPRs that hold 64-bit address. These instructions are used to access to VRAM, LDS and scratch buffer. The prefix is "flat".

Suffix (if present) means data type and size. Supported data types are indicated by the following suffixes:

i — signed integer;
u — unsigned integer;
f — floating-point;
b — binary (for bitwise operations).

The data type size can be 8, 16, 24, 32 and 64. Some instructions contain double suffix. For example, V_MUL_HI instruction family (V_MUL_HI_I32_I24, V_MUL_HI_U32_U24).

The rest of command name defines the operation. Some operations do not have direct equivalents in OpenCL. Such operations are decompiled to several OpenCL instructions. Otherwise, some assembly instructions are grouped and decompiled into a single OpenCL instruction.

AMD GCN devices do not have a call stack. Consequently, all the function calls are inlined into a kernel. Therefore, assembly code does not have any information about functions. We can only guess that there was a function if we discovered identical code fragments (ignore register renaming). But such an analysis is not considered in this paper.

### Search for memory accesses

Assembly instructions processing starts from searching for memory accesses. The basic data structure used in the following algorithms is called *Register*. It holds the information about a single register and contains the following fields: *version, type, integrity. Integrity* can hold one of these values: {*entire, high_part, low_part*}. *Entire* means the register holds the whole 32 (or less) bit variable. Other values mean the register holds a part of 64 bit variable.

AMD ABI documentation contains description for OpenCL work-item built-in functions.

At this stage, the following functions are supported:

get_global_id(uint dimindx);
get_global_offset(uint dimindx);
get_local_id(uint dimindx);
get_global_size(uint dimindx);
get_local_size(uint dimindx);
get_group_id(uint dimindx);
get_num_groups(uint dimindx);
get_work_dim().

The result of these functions is stored to specific addresses. Therefore, if such an address is loaded into a register, then further access to that register means a call to this function.

The get_global_id(dim) function returns a global thread identifier that is unique in the entire task space. dim can take possible values of 0, 1 or 2. Since the thread numbering can be shifted in kernels, in order to get a thread index starting from zero, there is the following idiom:

```
uint idx0 = get_global_id(0) –
get_global_offset(0);
```

This thread index is often used to refer to an array. We parse this index and get_global_id in the following steps:

In *the first step*, we detect get_global_offset(uint dimindx). The value of this function is stored in global memory by address s[4:5]. So instruction s_load_dwordx2 s[2:3] s[4:5] 0x0 means get_global_offset(0) stored in register pair s2, s3.

*The second step* is determining the local ID: get_local_id(0). Local ID is stored in register v0 before the program starts executing, and in case of 2D or 3D index range get_local_id(1) and get_local_id(2) are stored in v1 and v2, respectively. Therefore the field *type* of these registers data is filled before the instruction processing (it corresponds to get_local_id(uint dimindx)).

*The third step* is identifying the work-group ID: get_group_id(uint dimindx) . The result of calling this function is also compile-time constant and stored before the program execution. If "useargs" is used by the kernel in the configuration, get_group_id(0) is stored in register s6, and (in case of 2D and 3D index range) get_group_id(1) and get_group_id(2), are stored in s7 and s8, respectively. This instruction is processed like the previous one. The registers fields *type* are filled with corresponding values before the instruction processing.

*The fourth step* is discovering the work-group size. In OpenCL this value can be retrieved using get_local_size function. It is impossible to determine the call to this function from the assembly code. This is because the value of this function call is replaced by numeric constant. Therefore, we have no semantic information about this number in the assembly code. However, we have obtained work-group size in the previous section.

*The last step* is multiplying the work-group ID by the work-group size, and then the local thread ID.

Function get_global_id(uint dimindx)is deconstructed in similar way but with the addition of the offset value.

The result of a function that returns the size of the workspace in a given dimension, get_global_size(uint dimindx), is stored in global memory by address s[4:5]+ 0xc, 0x10 or 0x14 depending on the dimension. Processing of this instruction is same with get_global_offset: data type inference is done using instruction s_load_dword with offsets (0xc, 0x10, 0x14).

Next, consider a function that returns the number of work-groups that will run the kernel for a given dimension, get_num_groups(uint dimindx). The value is obtained by dividing the size of the workspace by the size of the work-group for a given dimension.

The last function to consider is get_work_dim(). It returns the number of dimensions used. The value is obtained when dword is loaded from the registers storing a pointer to kernel settings — s[4:5], with an offset of 0x20010. Processing of this instruction is the same with get_global_offset и get_global_size.

The result of matching with the presented templates is a restoration of work-item built-in functions.

Also, calls to array elements and simple arithmetic operations were supported.

## Search for control structures

The decompiler was implemented using an algorithm based on structural analysis [21]. At first step, we construct the control flow graph [22]. After that, we transform it to region graph. Initially, each instruction represents one region.

The analysis process in based on depth-first search. Each node is checked whether it is a header of one of known templates. If the template is determined, all the nodes corresponding to this template are merged into a single node. This process is iterated until the single node remains.

Our decompiler supports the if construction. The template presented for it in Fig. 2, corresponds to the one described in theoretical solutions, and does not require any additional transformations for detection and decompilation.

The region graph processing algorithm is illustrated by the example shown in Fig. 3. The algorithm consists of the following steps:

1. Regions ##1–3 are not beginning of any known templates. Region #4 in conjunction with regions #5 and #6 constitute an *if template*. However, region #6 is connected with another regions. So, we merge only regions #4 and #5 into a new region #7.

2. Regions #1 and #2 are not beginning of any known templates. Regions #3 and #7 constitute a *linear region*. Merge them into a new region #8.

3. Region #1 is not beginning of any known templates. Regions #2, #8 and #6 constitute an *if template*. Region #6 is connected with another region (region #1), so merge only regions #2 and #8 into a new region #9.

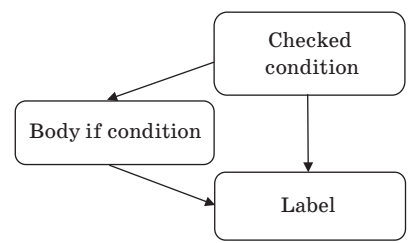

■ *Fig. 2.* Template for if statement

4. Regions #1, #9 and #6 constitute an *if template*. Merge them into a new region #10.

5. There is a single region now. So, we extracted all flow-control information from the region graph and can now generate OpenCL code.

The main difference with CPU *if-else template* is the presence of a 64-bit mask, which is responsible for the execution of threads. This is because 64 threads have the same instruction pointer. AMD compiler generates *if-else* construction in several forms. We demote the most frequent form as the *first form*. The *first form* is shown in Fig. 4, *a*. For more convenient processing, this template was reduced to the form shown in Fig. 4, b (*standard form*).

In this paper we also consider another two frequent forms. We denote them as the *second form* and the *third form*. The *second form* is shown in Fig. 5, *a*. The *third form* is shown in Fig. 5, *b*. The reduction the second form of *if-else template* to the *standard form* (see Fig. 4, *b*) consists of two steps:

1. Transformation to the *first form* of *if-else template*.

2. Reduction to *standard form*.

The second form of if-else template looks like the if template. But the main difference is the second change exec mask and else condition body before restoration of exec mask. The main difference between the first form and the second form is a quantity of "goto" labels.

The transformation of the *second form* to the *first form* is made by fake insertion of the second label after the else condition body and condition jump to the second label before it. The transformation of the *third form* is made similarly.

The processing of nested structures is the following. Firstly, the most nested structures are detected using control instruction templates. Detected structures are combined in the region graph. After that, the most nested of the remaining structures can be detected. The processing is continued until the root structure is combined in the region graph.

When processing branches, it was taken into account that at a vertex that has several ancestors, the values of registers can be determined ambiguously. And if in the future some of these registers were used, then variables were created for them. In the implementation, this was done by assigning versions to registers and working with them [23, 24].

The last considered in this paper control structure is the ternary operator. It is represented in the assembly code of one instruction and does not require overlapping templates.

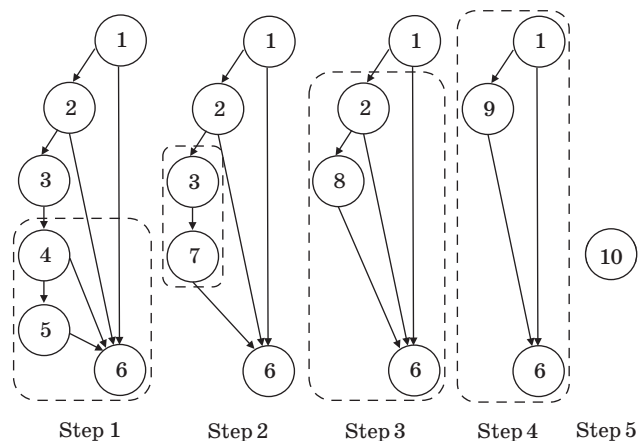

*Fig. 3.* Example of region graph handling

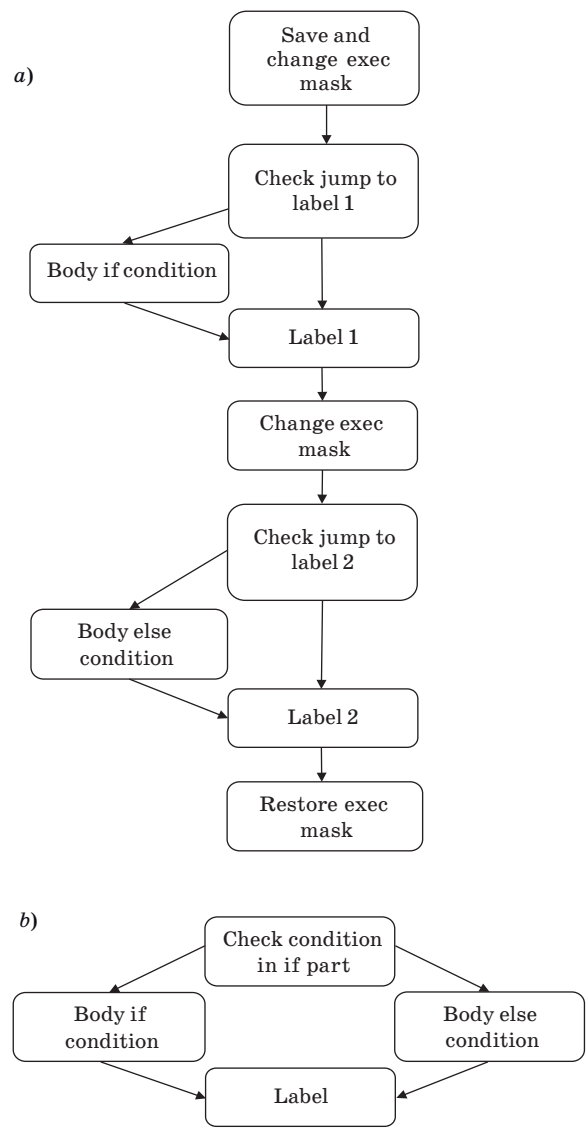

*Fig. 4.* Templates for if-else conditions part 1: *a* — with two labels; *b* — standard form

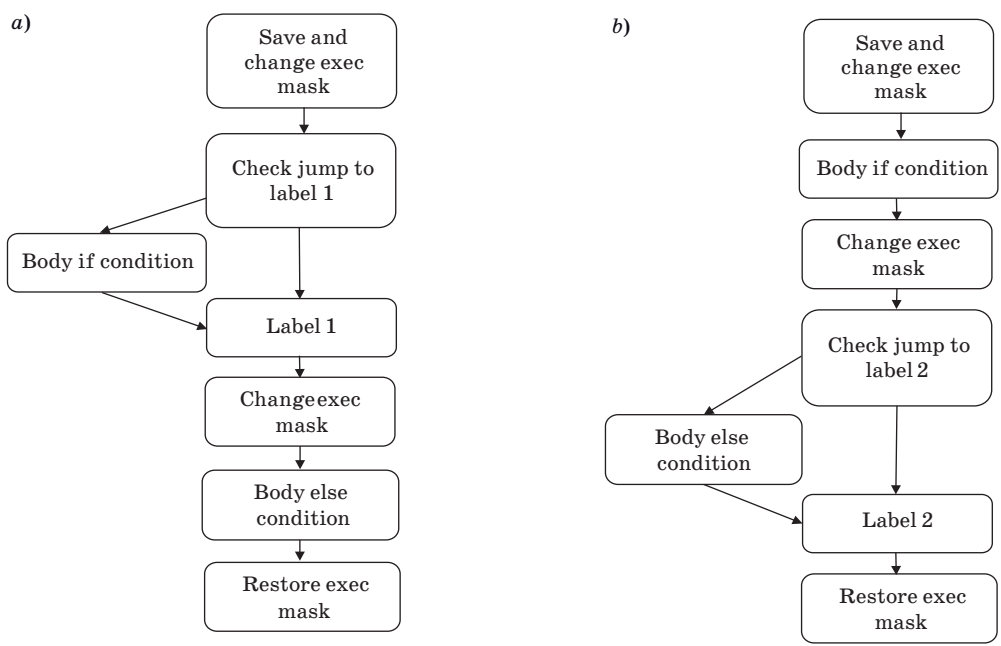

■ *Fig. 5.* Templates for if-else conditions part 2: *a* — with label in the if part; *b* — with label in the else part

## Data type recovery

Two ways of data type recovery were implemented: from the .config section of kernel and using assembly instructions. The .config section contains data types for kernel arguments. For example, the .config section of kernel with signature void copy(__global int *data, int x) is shown in Listing 2. As can be seen from Listing 2, data type for kernel arguments can be restored unambiguously.

*Listing 3.* Kernel arguments
```
.kernel copy
 .config
  .dims x
  .cws 64, 1, 1
  .sgprsnum 13
  .vgprsnum 3
  .floatmode 0xc0
  .pgmrsrc1 0x00ac0040
  .pgmrsrc2 0x0000008c
  .dx10clamp
  .ieeemode
  .useargs
  .priority 0
  .arg _.global_offset_0, "size_t", long
  .arg _.global_offset_1, "size_t", long
  .arg _.global_offset_2, "size_t", long
  .arg _.printf_buffer, "size_t", void*, global, , rdonly
  .arg _.vqueue_pointer,"size_t", long
  .arg _.aqlwrap_pointer,"size_t", long
  .arg data, "int*", int*, global,
  .arg x, "int", int
```

Data type determination using assembly instructions is based on instruction suffixes. For example, instruction

```
s_add_u32 s0, s4, s0
```

means sum of two unsigned 32-bit integers.

## Practical implementation

As a practical implementation of this research, the OpenCL Decompiler tool was developed. At this moment, it supports only a reduced set of AMD GCN ISA.

The OpenCL Decompiler was implemented in Python 3. It requires an assembly file compatible with CLRX Disassembler [25] output or CodeXL assembly listing as input data.

The output of the OpenCL Decompiler is a valid OpenCL file. All decompiled kernels can be compiled and executed on AMD GPUs. The exception is case when the decompiler gets an unsupported instruction. In this case decompiler lefts unsupported assembly code as is in inline assembly (inline assembly is not supported by AMDGPU-Pro driver and cannot be compiled).

The source code is available at https://github.com/sudo-team-company/OpenCLDecompiler.

The repository has about 931 synthetic tests and real free open-source kernels. Decompiler passes all the tests in the repository, which confirms correctness described functionality.

The examples of the real kernels are mask_kernel and weighted_sum_kernel (https://github.com/ganyc717/Darknet-On-OpenCL/blob/master/darknet_cl/cl_kernels/blas_kernels_1.cl). The result of their decompilation is in folder real_tests (https://github.com/sudo-team-company/OpenCLDecompiler/tree/master/tests/real_kernels).

These tests confirm the compliance of the theoretical considerations and practical results.

**Conclusion**

In this paper, a decompiling method for AMD GPU assembly was described. It has an implementation called *OpenCLDecompiler* and was introduced into Sudo Ltd. The *OpenCLDecompiler* tool was demonstrated on real open-source projects. All of this reveals the practical applicability of described method.

The described method is based on standard techniques for CPU decompilers but some techniques required significant modification for massive parallel architecture.

Decompiler works with any valid assembly code. However, restoration of some complicated loop constructions and some instructions is not implemented. In this case all supported assembly instructions are decompiled into a pseudo-code in accordance with their documentation. Unsupported instructions are remained unchanged. This approach does not provide full-fledged OpenCL code but significantly facilitate further manual code analysis.

It is further planned to extend the set of supported instructions and support the new RDNA architecture [26] and processing of more complicated flow control instructions.

**Financial support**

This work was supported by Sudo Ltd, project No cr-776, and National Center for Cognitive Research of ITMO University.

**References**

1. *OpenCL Overview — The Khronos Group Inc.* Available at: https://www.khronos.org/opencl (accessed 25 January 2020).
2. Jedel' G. E. *Parallel'nye vychislenija na graficheskih processorah Nvidia CUDA.* In: *Sbornik izbrannyh statej nauchnoj sessii TUSUR* [Parallel computing on GPU Nvidia Cuda. In: Collection of selected articles of the TUSUR scientific session]. Tomsk, Tomskij gosudarstvennyj universitet sistem upravlenija i radiojelektroniki, 2020, no. 1-1, pp. 41–43 (In Russian).
3. Kondakov E. V. Parallel computing on GPUs. *Materialy XL nauchno-prakticheskoj konferencii "Nauka XXI veka: problemy, poiski, reshenija"* [Materials of the XL Scientific-Practical Conference "Science of the 21st Century: Problems, Search, Solutions"]. Miass, 2016, pp. 34–39 (In Russian).
4. Pryadko S. A., Troshin A. Y., Kozlov V. D., Ivanov A. E. Parallel programming technologies on computer complexes. *Radio industry*, 2020, vol. 30, no. 3, pp. 28–33 (In Russian). doi:10.21778/2413-9599-2019-30-3-28-33
5. *The OpenCL Specification. Version: 1.2.* Available at: https://www.khronos.org/registry/OpenCL/specs/opencl-1.2.pdf (accessed 26 January 2020).
6. Yifan Sun, Trinayan Baruah, Saiful A. Mojumder, Shi Dong, Xiang Gong, Shane Treadway, Yuhui Bao, Spencer Hance, Carter McCardwell, Vincent Zhao, Harrison Barclay, Amir Kavyan Ziabari, Zhongliang Chen, Rafael Ubal, José L. Abellán, John Kim, Ajay Joshi, and David Kaeli. MGPUSim: Enabling Multi-GPU performance modeling and optimization. *The 46th Annual International Symposium on Computer Architecture (ISCA '19)*, June 22–26, 2019, Phoenix, AZ, USA. ACM, New York, NY, USA, 2019, 13 p. https://doi.org/10.1145/3307650.3322230
7. *ROCm — AMDGPU Compute Application Binary Interface.* Available at: https://github.com/ROCm-Developer-Tools/ROCm-ComputeABI-Doc/blob/master/AMDGPU-ABI.md (accessed 12 April 2020).
8. Mikhailov A. A., Khmelnov A. E. Control flow graph visualization. *BSU Bulletin. Mathematics, Informatics*, 2018, no. 2, pp. 50–62 (In Russian). doi:10.18101/2304-5728-2018-2-50-62/issn2304-5728
9. Klimenko V. Y., Saradzhishvili S. E. Optimization of loop search in flowgraphs. *Veles*, 2019, no. 10–1 (76), pp. 63–66 (In Russian).
10. Menshikov M. A. Effective translation of directed acyclic graphs to intermediate representation. *Processy upravlenija i ustojchivost'*, 2020, no. 1, pp. 271–275 (In Russian).
11. Jumaganov A. S. A combined method of similar code sequences search in executable files. *V mezhdunarodnaja konferencija i molodjozhnaja shkola «Informacionnye tehnologii i nanotehnologii»* [The V International Conference and Youth School "Information Technology and Nanotechnology"]. Samara, 2019, pp. 639–646 (In Russian).
12. Treshhev I. A, Serikov V. A. A practical approach to the implementation of the decompilation of machine code. *Sbornik materialov IV Vserossijskoj nauch-* *no-prakticheskoj konferencii (s mezhdunarodnym uchastiem) "Informacionnye tehnologii v jekonomike i upravlenii"* [Collection of Materials of the IV All-Russian Scientific-Practical Conference (with international participation) "Information Technologies in Economy and Management"]. Mahachkala, 2020, pp. 168–173 (In Russian).
13. Izrailov K. E. Applying of genetic algorithms to decompile machine code. *Zashhita informacii*, Insajd, Saint-Petersburg, 2020, no. 3(93), pp. 24–30 (In Russian).
14. Andreev A. A., Datsun N. N. Optimization analisys of fore language translation stages. *Materialy Vserossijskoj nauchno-prakticheskoj konferencii molodyh uchenyh s mezhdunarodnym uchastiem "Matematika i mezhdisciplinarnye issledovanija"* [Materials of the All-Russian Scientific and Practical Conference of Young Scientists with International Participation "Mathematics and Interdisciplinary Research"]. Perm', 2020, pp. 11–15 (In Russian).
15. Katz D. S., Ruchti J., and Schulte E. Using recurrent neural networks for decompilation. *IEEE 25th International Conference on Software Analysis, Evolution and Reengineering (SANER)*, 2018, pp. 346–356. doi: 10.1109/SANER.2018.8330222
16. Andrea Gussoni, Alessandro Di Federico, Pietro Fezzardi, and Giovanni Agosta. A comb for decompiled C code. *15th ACM Asia Conference on Computer and Communications Security (ASIA CCS'20)*, October 5–9, 2020, Taipei, Taiwan, ACM, New York, NY, USA, 15 p. https: //doi.org/10.1145/3320269.3384766
17. Gusenko M. The use of regular expressions for decompiling static data. *Software Systems and Computational Methods*, 2017, no. 2, pp. 1–13. doi:10.7256/2454-0714.2017.2.22608
18. Liu Z., and S. Wang. How far we have come: Testing decompilation correctness of C decompilers. *Proceedings of the 29th ACM SIGSOFT International Symposium on Software Testing and Analysis (ISSTA 2020)*, Association for Computing Machinery, New York, NY, USA, pp. 475–487. doi:10.1145/3395363.3397370
19. Ayoshin I. T. Reverse engineering of software by IDA Pro. *Aktual'nye problemy aviacii i kosmonavtiki*, 2018, vol. 3, no. 4 (14), pp. 808–809 (In Russian).
20. Vorob'ev A. M., Bocvin A. S., Nagibin D. V. Functional analysis of Ghidra — a framework for reverse engineering. *Metody i tehnicheskie sredstva obespechenija bezopasnosti informacii*, 2019, no. 28, pp. 86–88 (In Russian).
21. Derevenec E. O., Troshina E. N. Structural analysis in a decompilation problem. *Prikladnaya informatika*, 2009, no. 4(22), pp. 87–99 (In Russian).
22. Blank Ya. A., Savkin M. K. Control flow graph. *Materialy regional'noj nauchno-tehnicheskoj konferencii "Naukoemkie tehnologii v priboro- i mashinostroenii i razvitie innovacionnoj dejatel'nosti v vuze"* [Materials of the Regional Scientific and Technical Conference "Science-Intensive Technologies in Instrument and Mechanical Engineering and the Development of Innovative Activities in the University"]. Kaluga, 2016, pp. 75–78 (In Russian).
23. Masud A. N., and Ciccozzi F. More precise construction of static single assignment programs using reaching definitions. *Journal of Systems and Software*, 2020, vol. 166. doi:10.1016/j.jss.2020.110590
24. Masud A. N., and Ciccozzi F. Towards Constructing the SSA form using Reaching Definitions Over Dominance Frontiers. *19th International Working Conference on Source Code Analysis and Manipulation (SCAM)*, 2019, pp. 23–33. doi: 10.1109/SCAM.2019.00012
25. *CLRadeonExtender*. Available at: http://clrx.native-boinc.org (accessed 11 April 2020).
26. Secrets of the new RDNA graphics architecture revealed. *Otkrytye sistemy. SUBD*, 2019, no. 3, p. 5 (In Russian).

К. И. Михайленко[а,б], магистрант, младший программист, orcid.org/0000-0002-6168-2653, Kristina.Mihajlenko@gmail.com
М. А. Лукин[а,б], канд. техн. наук, технический директор, orcid.org/0000-0002-1088-3324, lukinma@gmail.com
А. С. Станкевич[а], канд. техн. наук, доцент, orcid.org/0000-0002-3532-8941, stankev@itmo.ru
[а]Национальный исследовательский университет ИТМО, Кронверкский пр., 49, Санкт-Петербург, 197101, РФ
[б]ООО «Судо», Нахимова ул., 20, Санкт-Петербург, 199226, РФ

**Введение:** декомпиляторы являются удобным инструментом для анализа и поддержки программ при отсутствии исходного кода. Существуют декомпиляторы для многих архитектур и языков программирования, но для графических процессоров семейств AMD GCN и RDNA такого инструмента в настоящее время нет. **Цель:** разработать декомпилятор ассемблерного кода AMD GPU в язык программирования OpenCL, широко используемый для программирования на устройствах класса GPGPU. **Результаты:** определены алгоритмы первичной обработки ассемблерного кода: выделение названия программы, параметров и тела программы; поиска обращений к данным и к элементам массивов; извлечения системных значений; поиска и декомпиляции некоторых арифметических операций. Также выработан метод восстановления типов и для работы с локальной памятью. Разработаны шаблоны для определения управляющих конструкций. **Практическая значимость:** предложенные алгоритмы и метод реализованы на языке Python в виде инструмента OpenCLDecompiler, поддерживающего достаточно большое подмножество команд архитектуры AMD GCN. Разработанный инструмент производит декомпиляцию ассемблерного кода, полученного в результате дизассемблирования исполняемого файла, в код на языке OpenCL, что позволяет сократить трудозатраты на анализ ассемблерного кода.

**Ключевые слова** — декомпилятор, дизассемблер, OpenCL, AMD GCN, GPGPU, граф потока управления, обратная разработка.

## Уважаемые авторы!

**При подготовке рукописей статей необходимо руководствоваться следующими рекомендациями.**

Статьи должны содержать изложение новых научных результатов. Название статьи должно быть кратким, но информативным. В названии недопустимо использование сокращений, кроме самых общепринятых (РАН, РФ, САПР и т. п.).

Текст рукописи должен быть оригинальным, а цитирование и самоцитирование корректно оформлено.

Объем статьи (текст, таблицы, иллюстрации и библиография) не должен превышать эквивалента в 20 страниц, напечатанных на бумаге формата А4 на одной стороне через 1,5 интервала Word шрифтом Times New Roman размером 13, поля не менее двух сантиметров.

Обязательными элементами оформления статьи являются: индекс УДК, заглавие, инициалы и фамилия автора (авторов), ученая степень, звание (при отсутствии — должность), полное название организации, аннотация и ключевые слова на русском и английском языках, ORCID и электронный адрес одного из авторов. При написании аннотации не используйте аббревиатур и не делайте ссылок на источники в списке литературы. Предоставляйте подрисуночные подписи и названия таблиц на русском и английском языках.

Статьи авторов, не имеющих ученой степени, рекомендуется публиковать в соавторстве с научным руководителем, наличие подписи научного руководителя на рукописи обязательно; в случае самостоятельной публикации обязательно предоставляйте заверенную по месту работы рекомендацию научного руководителя с указанием его фамилии, имени, отчества, места работы, должности, ученого звания, ученой степени.

**Формулы** набирайте в Word, не используя формульный редактор (Mathtype или Equation), при необходимости можно использовать формульный редактор; для набора одной формулы не используйте два редактора; при наборе формул в формульном редакторе знаки препинания, ограничивающие формулу, набирайте вместе с формулой; для установки размера шрифта никогда не пользуйтесь вкладкой Other…, используйте заводские установки редактора, не подгоняйте размер символов в формулах под размер шрифта в тексте статьи, не растягивайте и не сжимайте мышью формулы, вставленные в текст; в формулах не отделяйте пробелами знаки: + = −.

Для набора формул в Word никогда не используйте Конструктор (на верхней панели: «Работа с формулами» — «Конструктор»), так как этот ресурс предназначен только для внутреннего использования в Word и не поддерживается программами, предназначенными для изготовления оригинал-макета журнала.

При наборе символов в тексте помните, что символы, обозначаемые латинскими буквами, набираются светлым курсивом, русскими и греческими — светлым прямым, векторы и матрицы — прямым полужирным шрифтом.

**Иллюстрации** предоставляются отдельными исходными файлами, поддающимися редактированию:

— рисунки, графики, диаграммы, блок-схемы предоставляйте в виде отдельных исходных файлов, поддающихся редактированию, используя векторные программы: Visio (*.vsd, *.vsdx); Coreldraw (*.cdr); Excel (*.xls); Word (*.docx); Adobe Illustrator (*.ai); AutoCad (*.dxf); Matlab (*.ps, *.pdf или экспорт в формат *.ai);

— если редактор, в котором Вы изготавливаете рисунок, не позволяет сохранить в векторном формате, используйте функцию экспорта (только по отношению к исходному рисунку), например, в формат *.ai, *.esp, *.wmf, *.emf, *.svg;

— фото и растровые — в формате *.tif, *.png с максимальным разрешением (не менее 300 pixels/inch).

Наличие подрисуночных подписей и названий таблиц на русском и английском языках обязательно (желательно не повторяющих дословно комментарии к рисункам в тексте статьи).

**В редакцию предоставляются:**

— сведения об авторе (фамилия, имя, отчество, место работы, должность, ученое звание, учебное заведение и год его окончания, ученая степень и год защиты диссертации, область научных интересов, количество научных публикаций, домашний и служебный адреса и телефоны, e-mail), фото авторов: анфас, в темной одежде на белом фоне, должны быть видны плечи и грудь, высокая степень четкости изображения без теней и отблесков на лице, фото можно представить в электронном виде в формате *.tif, *.png с максимальным разрешением — не менее 300 pixels/inch при минимальном размере фото 40×55 мм;

— экспертное заключение.

**Список литературы** составляется по порядку ссылок в тексте и оформляется следующим образом:

— для книг и сборников — фамилия и инициалы авторов, полное название книги (сборника), город, издательство, год, общее количество страниц, doi;

— для журнальных статей — фамилия и инициалы авторов, полное название статьи, название журнала, год издания, номер журнала, номера страниц, doi;

— ссылки на иностранную литературу следует давать на языке оригинала без сокращений;

— при использовании web-материалов указывайте адрес сайта и дату обращения.

Список литературы оформляйте двумя отдельными блоками по образцам lit.dot на сайте журнала (http://i-us.ru/paperrules): Литература и References.

Более подробно правила подготовки текста с образцами изложены на нашем сайте в разделе «Руководство для авторов».

## Контакты

Куда: 190000, Санкт-Петербург,
Б. Морская ул., д. 67, ГУАП, РИЦ
Кому: Редакция журнала «Информационно-управляющие системы»
Тел.: (812) 494-70-02
Эл. почта: ius.spb@gmail.com
Сайт: www.i-us.ru